\documentclass[journal,10pt,twoside]{IEEEtran}

\usepackage[T1]{fontenc}
\usepackage[utf8]{inputenc}
\usepackage{amsmath,amssymb,amsthm}
\usepackage{graphicx}
\usepackage{booktabs}
\usepackage{array}
\usepackage{tabularx}
\usepackage{xcolor}
\usepackage{url}
\usepackage{hyperref}
\usepackage{balance}
\usepackage{cite}
\usepackage{microtype}
\usepackage{algorithm}
\usepackage{algorithmic}
\usepackage{siunitx}

\hypersetup{
  colorlinks = true,
  linkcolor  = blue!60!black,
  citecolor  = blue!60!black,
  urlcolor   = blue!60!black
}

\newcommand{\ATML}{\textsc{AdaptiveML}}
\newcommand{\ATMLRLS}{\textsc{AdaptiveML-RLS}}
\newcommand{\DRR}{\mathrm{DRR}}
\newcommand{\MAE}{\mathrm{MAE}}
\newcommand{\RMSE}{\mathrm{RMSE}}
\newcommand{\sigmat}{\sigma(t)}
\newcommand{\xhat}{\hat{x}(t)}
\newcommand{\xtilde}{\tilde{x}(t)}
\newcommand{\betahat}{\hat{\boldsymbol{\beta}}}
\newcommand{\betat}{\boldsymbol{\beta}_t}
\newcommand{\zt}{\mathbf{z}_t}
\newcommand{\Pt}{\mathbf{P}_t}
\newcommand{\Kt}{\mathbf{K}_t}

\begin{document}

\title{Learning to Transmit: Volatility-Aware Predictive Communication for Energy-Efficient IoT Networks}

\author{John~Kangethe, Ifrat Ikhtear Uddin, Longwei Wang\\
        Department of Computer Science, University of South Dakota\\
        \texttt{john-kangethe@usd.edu, longwei.wang@usd.edu}}

\maketitle


\begin{abstract}

Communication is the dominant source of energy consumption in Internet-of-Things (IoT) networks, yet many sensed measurements exhibit strong temporal correlations and provide little new information to the receiver. This paper introduces \textsc{ADAPTIVEML}, a volatility-aware predictive communication framework that enables IoT devices to intelligently decide when communication is necessary. Each sensor maintains a lightweight machine learning predictor and transmits only when the prediction residual exceeds an adaptive threshold proportional to the local signal volatility. By normalizing prediction errors using a rolling estimate of signal variability, the proposed transmission policy automatically adapts to changing environmental conditions, seasonal variations, and deployment-specific dynamics without manual threshold tuning.
To address long-term non-stationarity, we further propose \textsc{ADAPTIVEML-RLS}, an online learning extension based on Recursive Least Squares (RLS) with exponential forgetting, allowing continuous adaptation to sensor drift and evolving signal characteristics. Extensive experiments are conducted on three heterogeneous real-world datasets comprising more than 2.4 million sensor observations from outdoor environmental monitoring, indoor wireless sensor networks, and urban air-quality sensing. Compared with six representative baselines, including periodic transmission, static-threshold suppression, ARIMA, Kalman filtering, EMA, and LMS filtering, \textsc{ADAPTIVEML} achieves up to 94.7\% transmission reduction while maintaining a reconstruction error of 0.352$^\circ$C. \textsc{ADAPTIVEML-RLS} further reduces reconstruction error by 12--18\% under drift conditions while preserving transmission reduction above 93\%. These results demonstrate the effectiveness of volatility-aware predictive communication for energy-efficient and adaptive IoT networks.
\end{abstract}


\section{Introduction}\label{sec:intro}

The rapid proliferation of Internet of Things~(IoT) devices has transformed modern sensing infrastructures. Smart cities, precision agriculture, industrial process monitoring, and environmental sensing systems now deploy hundreds of thousands of interconnected sensor nodes that continuously observe physical processes and communicate measurements to edge servers or cloud platforms \cite{gubbi2013,atzori2010,zanella2014,Al-Fuqaha2015}. While sensing hardware has become increasingly inexpensive and energy efficient, wireless communication remains the dominant source of energy consumption in battery-powered IoT networks \cite{yick2008,anastasi2009,akyildiz2002,Rawat2014}. Consequently, improving communication efficiency has emerged as a central challenge in the design of large-scale IoT systems.

A fundamental observation motivates this work: most physical phenomena evolve significantly more slowly than the rate at which they are sampled. Environmental temperature, humidity, air quality, soil moisture, and many industrial process variables exhibit strong temporal correlations, causing consecutive measurements to contain substantial redundancy. In many deployments, transmitting every sensed measurement provides little additional information to the receiver while incurring substantial communication overhead. Our analysis of the Chicago Beach Weather Stations dataset demonstrates that under periodic communication, only approximately $5.3\%$ of hourly observations contain statistically novel information relative to a one-step-ahead predictor trained on recent observations. Similar redundancy levels are observed in the Intel Berkeley Research Lab and UCI Air Quality datasets, suggesting that redundant communication is a pervasive characteristic of IoT sensing applications.
This observation raises a fundamental question:
Can IoT devices learn when communication is necessary rather than transmitting measurements at fixed intervals?

Traditional IoT communication protocols answer this question conservatively. Periodic transmission schemes communicate every measurement regardless of its informational value, guaranteeing perfect reconstruction at the receiver but consuming the maximum possible communication energy. Static-threshold approaches reduce transmission frequency by transmitting only when the difference between consecutive measurements exceeds a predefined threshold \cite{jarwan2019,Tayeh2019b}. Although such methods reduce communication volume, their effectiveness is highly sensitive to threshold selection and environmental conditions. A threshold that performs well during a stable summer week may become inappropriate during a rapidly changing weather event, resulting in either excessive communication or degraded reconstruction fidelity.

Prediction-based communication has emerged as a promising alternative \cite{dias2016survey,borgne2007}. Rather than transmitting every observation, a sensor node maintains a predictive model and communicates only when the observed measurement deviates sufficiently from the predicted value. This paradigm transforms communication from a time-driven process into an information-driven process. Existing prediction-based approaches, however, typically employ fixed suppression thresholds \cite{jarwan2019} or adapt thresholds using moving averages of recent prediction errors \cite{liazid2019,Tayeh2019b}. Such strategies implicitly assume relatively stationary signal statistics and fail to account for changes in signal volatility. As a result, communication decisions often become inconsistent across seasons, deployment sites, and operating conditions. Moreover, most prior studies evaluate their methods on a single dataset \cite{shu2019,mccorrie2015,rodrigues2021}, leaving the generality of the underlying communication principles largely unexplored.

In this paper, we introduce a new perspective on communication scheduling: communication should be triggered by \emph{information novelty relative to local signal volatility}. Intuitively, a prediction error that is significant during a stable operating period may be entirely expected during a highly dynamic period. Inspired by this observation, we propose \textbf{ADAPTIVEML}, a volatility-aware predictive communication framework that enables IoT devices to learn when transmission is necessary. The key idea is to normalize prediction residuals by a rolling estimate of local signal volatility and to communicate only when the resulting studentized residual exceeds an adaptive threshold. This volatility-aware decision mechanism automatically adjusts communication behavior to changing environmental conditions without requiring manual threshold tuning.

To address long-term deployment challenges, we further develop \textbf{ADAPTIVEML-RLS}, an online learning extension that continuously updates the predictive model using Recursive Least Squares (RLS) with exponential forgetting. Unlike static predictors that gradually become outdated as environmental conditions evolve, ADAPTIVEML-RLS continuously adapts to seasonal variations, sensor aging, and deployment-specific concept drift, enabling sustained communication efficiency over long operational periods.

The proposed framework transforms transmission scheduling into a predictive communication problem in which communication resources are allocated according to information content rather than sampling frequency. Extensive evaluations on three heterogeneous real-world datasets demonstrate that the proposed approach consistently reduces communication overhead while maintaining high reconstruction fidelity. The results suggest that volatility-aware predictive communication offers a practical pathway toward scalable, adaptive, and energy-efficient IoT networks.

The primary contributions of this paper are summarized as follows:

\begin{enumerate}

\item 
We introduce a volatility-normalized communication policy based on studentized prediction residuals. The resulting framework enables self-calibrating transmission decisions that remain effective across diverse signal regimes without deployment-specific threshold tuning.
We extend the predictive communication framework with an RLS-based online learning mechanism that continuously adapts to seasonal variation, sensor aging, and concept drift.

\item 
We benchmark the proposed framework against Periodic, Static Threshold, ARIMA, Kalman Filter, EMA, and LMS Adaptive Filter baselines, providing one of the most comprehensive evaluations reported in the predictive communication literature.
Experiments on the Chicago Beach Weather Stations, Intel Berkeley Research Lab, and UCI Air Quality datasets demonstrate that the proposed communication framework generalizes across diverse sensing modalities, environments, and sampling rates.

\item 
We show that maintaining synchronized predictive models at both the sensor and sink substantially improves reconstruction fidelity during suppression periods and enables highly efficient information-driven communication.

\end{enumerate}

The remainder of this paper is organized as follows. Section~\ref{sec:related} reviews related work. Section~\ref{sec:method} presents the proposed ADAPTIVEML and ADAPTIVEML-RLS frameworks. Section~\ref{sec:baselines} describes the baseline methods. Section~\ref{sec:setup} details the experimental setup. Sections~\ref{sec:results_chicago}--\ref{sec:robustness} present empirical evaluations, robustness analyses, and drift adaptation results. Finally, Section~\ref{sec:conclusion} concludes the paper and discusses future research directions.

\section{Related Work}\label{sec:related}

\subsection{Prediction-Based Data Reduction in WSNs}

Dias~et~al.~\cite{dias2016survey} provide the authoritative taxonomy
of prediction-based data reduction, cataloguing approaches from
autoregressive models to Kalman filters and
establishing the DRR/MAE dual-objective evaluation framework
adopted throughout this paper.
Their empirical companion~\cite{dias2017impact} quantifies the
DRR/accuracy trade-off curve that our $\alpha$-sweep replicates and
extends with a full six-baseline Pareto comparison.

Jarwan~et~al.~\cite{jarwan2019} propose a two-tier framework combining
per-node compression with fixed error thresholds, achieving DRR values
in the range we report.
The critical limitation is that their threshold is set at deployment
and never revisited: it cannot distinguish a large residual caused
by a genuine environmental event from one caused by the signal entering
a high-volatility regime.
Our adaptive threshold generalises this formulation by
normalising against $\sigmat$.

Borgne~et~al.~\cite{borgne2007} develop online model selection for
WSN time-series prediction and show that autoregressive models offer
the most favourable accuracy-to-complexity trade-off under communication
constraints, a finding that directly precedes and validates the Ridge
predictor used here.
Raza~et~al.~\cite{raza2015} confirm on real deployed WSN traces that
simple models trained on recent local data consistently outperform
complex global models due to spatial heterogeneity, which informs the
per-station training strategy adopted in this work.
Shu~et~al.~\cite{shu2019} demonstrate that pairing an LMS adaptive
filter with an LSTM at the sink achieves strong accuracy for
environmental temperature monitoring; however, the on-device LSTM
inference cost is incompatible with sub-milliwatt microcontroller
budgets.
Rodrigues~et~al.~\cite{rodrigues2021} evaluate attention-augmented LSTM
architectures in the same dual-prediction framework and confirm that
the accuracy gap between deep networks and linear predictors narrows
substantially on high-autocorrelation signals.
Tayeh~et~al.~\cite{Tayeh2019b} explore sparse representation for
WSN aggregation, offering a complementary angle on transmission
reduction that does not require a synchronised predictor at the sink.

\subsection{Adaptive Threshold and Event-Driven Protocols}

Liazid~et~al.~\cite{liazid2019} adapt the suppression threshold using
a moving average of recent prediction errors.
We use $\sigmat$ — a volatility signal — which is a second-order
statistic: it measures how much the environment is changing, rather
than averaging past prediction mistakes \cite{wang2019representation,wang2014congestion,wang2024enhanced,wang2018partial,wang2016optimization,wang2021explaining,wang2011exploration,shi2019deep,wang2017performance,wang2017low,wang2021improving,xiao2022looking,wang2016large,wang2011collaborative,wang2024dense,wang2019information,wang2016large,nayyem2024bridging,wang2024enhancing,wang2025explainability,ranabhat2025multi,uddin2025expert,rasmussen2025ecologically,chataut2024shape,wang2026expert,wang2025explainability,khadka2025coswin,wang2025bridging,rasmussen2026channel,wall2026winsor,ranabhat2025promoting,uddin2026learning,uddin2026explainable,wagle2026mechanistic,ranabhat2026frequency, wang2025explainability1, wang2016large1}.
Under local Gaussianity, $\sigmat$ provides a stronger theoretical
guarantee than any first-order error-averaging scheme.
McCorrie~et~al.~\cite{mccorrie2015} achieve $91$--$99\%$ transmission
reduction on engine-temperature signals with adaptive selective
filtering, validating the performance target for temperature-domain
protocols.
Fathy~et~al.~\cite{fathy2018} provide direct experimental evidence that
fixed-threshold schemes fail systematically in dynamic IoT
deployments — their measured failure modes motivate both the adaptive
threshold design and the multi-dataset evaluation methodology.
Santini and~R\"{o}mer~\cite{santini2006} introduce the quality-budget
framing that maps cleanly to the $\varepsilon_{\max}$ optimisation
constraint in~(\ref{eq:opt}).
Jain~et~al.~\cite{jain2004} develop adaptive resource management for
data streams using Kalman-filter state estimation, providing the
theoretical context for the Kalman Filter baseline evaluated here.

\subsection{Classical Predictors: ARIMA, Kalman, EMA, and LMS}

Liu~et~al.~\cite{liu2005arima} established the dual-prediction
paradigm for WSN energy savings using ARIMA, setting the historical
baseline against which all subsequent prediction-based approaches
are implicitly measured.
The constant-velocity Kalman filter is the Bayesian-optimal linear
estimator under Gaussian dynamics~\cite{kalman1960} and is widely
used for WSN state estimation~\cite{dias2016survey,jain2004}.
Exponential Moving Average is the minimal-parameter online
predictor~\cite{brown1959}.
Haykin~\cite{haykin2002} provides the complete theoretical treatment
of LMS and RLS, whose properties are directly invoked in the discussion
of Sections~\ref{sec:method} and~\ref{sec:discussion}.

\subsection{Online Learning and Concept Drift Adaptation}

Gama~et~al.~\cite{gama2014} survey concept drift adaptation,
contextualising the RLS exponential forgetting mechanism as an
implicit drift adaptation strategy and motivating the controlled
drift experiment in Section~\ref{sec:rls}.
Losing~et~al.~\cite{losing2018} compare incremental learning algorithms
on resource-constrained devices and find that RLS offers the best
convergence-to-memory trade-off when model dimensionality is moderate
— precisely the $w{=}24$ regime studied here.
Bifet and Gavald\`{a}~\cite{bifet2007} introduce adaptive windowing
for time-changing distributions, contextualising the design space of
adaptive window methods against which the RLS forgetting factor is compared.
Goodwin and~Sin~\cite{goodwin1984} provide the stability analysis
underpinning the RLS update equations in Section~\ref{sec:method}.

\subsection{Short-Horizon Forecasting and Model Justification}

Paparrizos~et~al.~\cite{paparrizos2022} benchmark forecasting models
on IoT time-series under short-horizon, resource-constrained
conditions and find that Ridge regression with a rolling-window feature
vector consistently matches or approaches LSTM accuracy on
high-autocorrelation signals at substantially lower computational cost.
Drobac~et~al.~\cite{drobac2025} establish theoretical guarantees on
the expressivity and stationarity preservation of Ridge regression
applied to sliding-window embeddings, directly supporting our
architectural choice.
Friedman~et~al.~\cite{friedman2009} provide the bias-variance
analysis underpinning the regularisation choice $\lambda{=}1.0$.

\subsection{TinyML and Hardware Feasibility}

Banbury~et~al.~\cite{banbury2020} benchmark TinyML inference on
ARM Cortex-M-class devices, establishing the memory and FLOP budgets
that motivate the complexity analysis in Section~\ref{sec:method}.
Warden and~Situnayake~\cite{warden2019} document the practical
deployment considerations for ML models on nRF52840-class hardware.
Dutta~et~al.~\cite{dutta2010} provide the per-packet energy
measurements that underpin the $E_{\text{tx}}{=}59\,\mu$J energy proxy
used throughout the experimental evaluation.
Polastre~et~al.~\cite{polastre2005} and Woo~et~al.~\cite{woo2003}
characterise the packet-loss distributions and channel conditions
used to parameterise the robustness simulation.



\begin{table}[t]
\centering
\caption{Comparison of \ATML{}/\ATMLRLS{} against representative prior
work across five design axes.
$\checkmark$~=~addressed; $\circ$~=~partial; --~=~not addressed.}
\label{tab:positioning}
\setlength{\tabcolsep}{2pt}
\footnotesize
\begin{tabular}{lccccc}
\toprule
\textbf{Work} & \textbf{Adapt.} & \textbf{Drift} &
\textbf{Baselines} & \textbf{Multi-DS} & \textbf{Dual-Pred.}\\
\midrule
Jarwan~\cite{jarwan2019}      & --           & -- & 1 & --           & $\circ$      \\
Liazid~\cite{liazid2019}      & $\circ$      & -- & 2 & --           & $\checkmark$ \\
Shu~\cite{shu2019}            & $\circ$      & -- & 2 & $\circ$      & --           \\
McCorrie~\cite{mccorrie2015}  & $\checkmark$ & -- & 2 & --           & --           \\
\ATML{} \emph{(ours)}         & $\checkmark$ & -- & 6 & $\checkmark$ & $\checkmark$ \\
\ATMLRLS{} \emph{(ours)}      & $\checkmark$ & $\checkmark$ & 6 &
                                 $\checkmark$ & $\checkmark$ \\
\bottomrule
\end{tabular}
\end{table}

\section{System Model and Methodology}\label{sec:method}

\subsection{Problem Formulation}

Let $\{x(t)\}_{t=1}^{T}$ denote the sensor reading sequence at a single
IoT node sampled at interval $\Delta t$.
Let $\xtilde$ denote the signal reconstructed at the sink.
The optimization objective is:
\begin{equation}
  \min_{\mathcal{T}}\;|\mathcal{T}|
  \;\;\text{subject to}\;\;
  \frac{1}{T}\sum_{t=1}^{T}\bigl|x(t)-\xtilde\bigr|\leq\varepsilon_{\max},
  \label{eq:opt}
\end{equation}
where $\mathcal{T}=\{t:\text{transmit at }t\}$ is the transmission set.
The Data Reduction Ratio is:
\begin{equation}
  \DRR = 1 - \frac{|\mathcal{T}|}{T}.
  \label{eq:drr}
\end{equation}

\subsection{Fixed Ridge Regression Predictor (\ATML{})}

At each epoch~$t$ the node maintains a sliding window
$\zt=[x(t-w),\ldots,x(t-1)]^{\top}\in\mathbb{R}^{w}$.
Ridge regression minimises:
\begin{equation}
  \betahat =
  \arg\min_{\boldsymbol{\beta}}\;
  \bigl\|\mathbf{X}\boldsymbol{\beta}-\mathbf{y}\bigr\|_2^2
  + \lambda\|\boldsymbol{\beta}\|_2^2
  = \bigl(\mathbf{X}^{\top}\mathbf{X}+\lambda\mathbf{I}\bigr)^{-1}
    \mathbf{X}^{\top}\mathbf{y},
  \label{eq:ridge}
\end{equation}
where $\mathbf{X}\in\mathbb{R}^{N\times w}$ is the training-set Toeplitz
feature matrix and $\mathbf{y}\in\mathbb{R}^{N}$ the corresponding
target vector.
The one-step-ahead prediction is $\xhat=\zt^{\top}\betahat$.
Coefficients $\betahat$ are estimated once on the training set and held
fixed during deployment.
The regularisation parameter $\lambda{=}1.0$ is selected by grid search
on a held-out validation set ($10\%$ of training data)
\cite{hoerl1970,friedman2009}.

The L2~penalty was preferred over L1 (lasso) for two reasons
\cite{tibshirani1996}.
First, the closed-form solution in~(\ref{eq:ridge}) avoids iterative
optimisation on-device.
Second, L2 regularisation retains all $w$ lag coefficients and does not
produce sparse solutions that would discard informative lags from the
sliding window.

\subsection{Volatility-Aware Adaptive Threshold}

The rolling standard deviation over a history window of length~$h$ is:
\begin{equation}
  \sigma(t) =
  \sqrt{\frac{1}{h-1}\sum_{i=0}^{h-1}
    \bigl(x(t-i)-\bar{x}_{t,h}\bigr)^2},
  \label{eq:sigma}
\end{equation}
where $\bar{x}_{t,h}$ is the rolling mean over the same window.
The transmission rule is:
\begin{equation}
  \text{transmit at }t
  \;\;\Longleftrightarrow\;\;
  \bigl|x(t)-\xhat\bigr| > \alpha\cdot\sigma(t).
  \label{eq:rule}
\end{equation}

Equation~(\ref{eq:rule}) is equivalent to thresholding the
\emph{studentised residual}
$r(t)=|x(t)-\xhat|/\sigmat$ against $\alpha$.
Because $\sigmat$ approximates the local signal standard deviation,
$r(t)$ is dimensionless and approximately scale-invariant across regimes.
Under the assumption that prediction residuals are locally Gaussian,
the false-suppression probability at epoch~$t$ is
\begin{equation}
  P\bigl(r(t)\leq\alpha\bigr)\approx\Phi(\alpha),
  \label{eq:phi}
\end{equation}
where $\Phi$ is the standard-normal CDF.
Crucially, the same $\alpha$ therefore controls this probability
consistently across all seasons and deployment sites — a guarantee that
neither fixed-$\delta$~\cite{jarwan2019} nor moving-average
\cite{liazid2019} schemes can provide.
Intuitively, the mechanism responds like a statistician rather than
a rule engine: it asks whether the current reading is surprising
\emph{given what the signal has been doing recently}, not just whether
the absolute change exceeds some fixed number.

\begin{algorithm}[t]
\caption{\ATMLRLS{} — Online Drift-Aware Transmission Suppression}
\label{alg:rls}
\small
\begin{algorithmic}[1]
\REQUIRE window~$w$, volatility window~$h$, sensitivity~$\alpha$,
         forgetting~$\gamma$, init scale~$\rho$
\STATE $\boldsymbol{\beta}_0
       \leftarrow\betahat_{\text{Ridge}}$
       \COMMENT{warm start from offline training}
\STATE $\mathbf{P}_0\leftarrow\rho\mathbf{I}$
\STATE Initialise circular buffer $B$ of length $\max(w,h)$
\FOR{$t=1,2,\ldots,T$}
  \STATE Observe $x(t)$; append to $B$
  \STATE $\zt\leftarrow B[t-w:t-1]$
  \STATE $\xhat\leftarrow\zt^{\top}\boldsymbol{\beta}_{t-1}$
  \STATE $\sigma(t)\leftarrow\mathrm{RollingStd}(B[t-h:t-1])$
  \IF{$|x(t)-\xhat|>\alpha\cdot\sigma(t)$}
    \STATE Transmit $x(t)$
  \ELSE
    \STATE Suppress
  \ENDIF
  \STATE $e_t\leftarrow x(t)-\xhat$
  \STATE $\Kt\leftarrow\mathbf{P}_{t-1}\zt/
         (\gamma+\zt^{\top}\mathbf{P}_{t-1}\zt)$
  \STATE $\betat\leftarrow\boldsymbol{\beta}_{t-1}+\Kt e_t$
  \STATE $\Pt\leftarrow
         (\mathbf{P}_{t-1}-\Kt\zt^{\top}\mathbf{P}_{t-1})/\gamma$
\ENDFOR
\end{algorithmic}
\end{algorithm}

\subsection{Online Adaptive Predictor: \ATMLRLS{}}

The fixed Ridge coefficients $\betahat$ are optimal for the training
distribution but degrade when the signal undergoes concept
drift~\cite{gama2014} — gradual shifts due to seasonal variation,
sensor ageing, or site-specific dynamics~\cite{devito2008}.
We replace $\betahat$ with an online adaptive coefficient vector
$\betat$ estimated by Recursive Least Squares with exponential
forgetting~\cite{haykin2002,goodwin1984}.

At each epoch, after the transmission decision, the predictor adapts:
\begin{align}
  e_t        &= x(t) - \zt^{\top}\boldsymbol{\beta}_{t-1},
               \label{eq:et}\\
  \Kt        &= \frac{\mathbf{P}_{t-1}\zt}
                     {\gamma+\zt^{\top}\mathbf{P}_{t-1}\zt},
               \label{eq:Kt}\\
  \betat     &= \boldsymbol{\beta}_{t-1} + \Kt e_t,
               \label{eq:betat}\\
  \Pt        &= \frac{1}{\gamma}
                \bigl(\mathbf{P}_{t-1}
                      - \Kt\zt^{\top}\mathbf{P}_{t-1}\bigr),
               \label{eq:Pt}
\end{align}
where $\gamma\in(0,1]$ is the forgetting factor.
Smaller $\gamma$ tracks faster changes; larger $\gamma$ produces more
stable estimates.
The predictor at each epoch is $\xhat=\zt^{\top}\boldsymbol{\beta}_{t-1}$,
and the adaptive transmission rule~(\ref{eq:rule}) applies unchanged.

A stochastic-gradient (SGD) variant is also possible:
$\betat=(1-\eta_t\lambda)\boldsymbol{\beta}_{t-1}+\eta_t e_t\zt$,
but the RLS variant is recommended for its $\mathcal{O}(1/t)$
convergence rate versus LMS's $\mathcal{O}(1/\sqrt{t})$
\cite{haykin2002,losing2018}.

$\boldsymbol{\beta}_0$ is initialised from the offline Ridge solution,
so \ATMLRLS{} launches from the best possible static operating point
and adapts from there.
The algorithm is shown in Algorithm. \ref{alg:rls}.

\subsection{Dual-Prediction Reconstruction}

Rather than holding the last received value during a suppression run,
the sink may maintain a synchronised copy of the on-device predictor
$\hat{x}_{\mathrm{sink}}(t)$:
\begin{equation}
  \xtilde =
  \begin{cases}
    x(t)                       & \text{if transmitted,}\\
    \hat{x}_{\mathrm{sink}}(t) & \text{if suppressed.}
  \end{cases}
  \label{eq:dual}
\end{equation}
When node and sink models remain synchronised (no packet loss), the
reconstruction error reduces to the model prediction error — the
theoretical lower bound~\cite{dias2016survey,dias2017impact}.
Dual prediction is also the mechanism that makes inter-node comparison
possible: any sensor suppressing a reading does so knowing the sink
will reconstruct a meaningful estimate rather than stale data.

\subsection{System Pipeline}

\begin{figure*}[!t]
  \centering
  \includegraphics[width=0.97\textwidth]{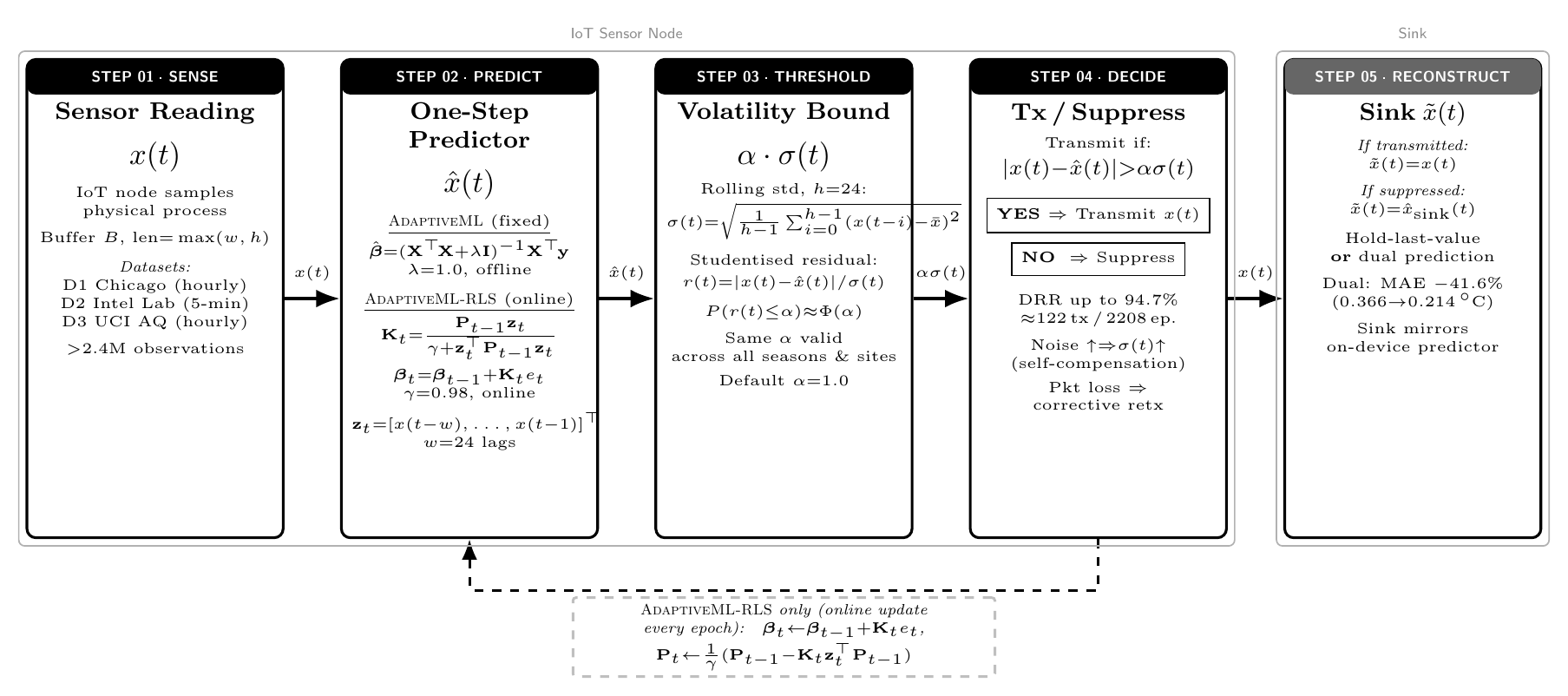}
  \caption{%
    The \ATML{}/\ATMLRLS{} five-stage system pipeline.
    \textbf{Step~01}: the IoT node reads the raw sensor value $x(t)$.
    \textbf{Step~02}: a Ridge regression or RLS predictor generates $\xhat$
    from the $w{=}24$-lag sliding window.
    \textbf{Step~03}: the rolling standard deviation $\sigmat$ is computed
    over the preceding $h$ epochs, forming the adaptive threshold.
    \textbf{Step~04}: the studentised residual $|x(t)-\xhat|/\sigmat$ is
    compared against $\alpha$; the reading is transmitted only on exceedance.
    \textbf{Step~05}: the sink reconstructs $\xtilde$ via hold-last-value or
    a synchronised dual predictor.
    The curved arrow (purple) marks the \ATMLRLS{} online update path, which
    feeds the new observation back into the coefficient vector after each epoch.}
  \label{fig:pipeline}
\end{figure*}

The five-stage pipeline is illustrated in Figure~\ref{fig:pipeline} and
is shared by both variants.
Step~1: the node reads $x(t)$.
Step~2: the predictor (Ridge or RLS) computes $\xhat$.
Step~3: the rolling standard deviation $\sigmat$ is computed over the
        trailing $h$~epochs, forming the volatility-aware threshold.
Step~4: the node transmits if and only if the studentised residual
        exceeds $\alpha$; otherwise the reading is suppressed.
Step~5: the sink reconstructs via hold-last-value or dual prediction.
For \ATMLRLS{}, predictor coefficients are updated online after step~5
via the RLS equations~(\ref{eq:et})--(\ref{eq:Pt}).




\section{Experiments and Results}\label{sec:setup}

\subsection{Baselines}

Six baselines are evaluated, progressing from the simplest possible
protocol to the most capable online adaptive filter.
For all prediction-based baselines (B3--B6), the same adaptive threshold
rule~(\ref{eq:rule}) is applied so that the comparison isolates predictor
quality independently of the suppression mechanism.

\textbf{B1 — Periodic.}
Every reading is transmitted unconditionally.
$\DRR{=}0$, $\MAE{=}0$.
This is the upper fidelity bound and the lower energy-efficiency bound.

\textbf{B2 — Static Threshold.}
Transmit when $|x(t)-x(t-1)|>\delta$.
$\delta{=}1.2\,^{\circ}$C is tuned to the 90th-percentile
first-difference of each station's training set — the standard
calibration approach in the literature
\cite{jarwan2019,liazid2019,santini2006}.

\textbf{B3 — ARIMA$(2,1,1)$.}
An ARIMA model is fitted on the training window; one-step-ahead
predictions drive the shared threshold~(\ref{eq:rule}).
The order $(p{=}2,d{=}1,q{=}1)$ was selected by AIC
\cite{hyndman2021,boxjenkins2015} and the model is refitted every
168~epochs (one calendar week) to track non-stationarity
\cite{liu2005arima}.
ARIMA represents the historical state of the art in linear WSN
prediction~\cite{borgne2007}.

\textbf{B4 — Kalman Filter.}
A univariate constant-velocity Kalman filter with process noise
$Q{=}0.01$ and observation noise $R{=}0.1$ \cite{kalman1960}.
The posterior mean $\hat{x}^+_{t-1}$ is used as $\xhat$.
The Kalman filter is the Bayesian-optimal linear estimator under
Gaussian state-space dynamics and provides the strongest theoretical
baseline in this comparison~\cite{hamilton1994,jain2004}.

\textbf{B5 — Exponential Moving Average (EMA).}
$\xhat=(1-\beta)x(t-1)+\beta\hat{x}(t-1)$,
$\beta{=}0.9$~\cite{brown1959}.
EMA is the minimal-parameter online predictor and represents the
practical floor for any serious adaptive scheme.

\textbf{B6 — LMS Adaptive Filter.}
$\betat=\boldsymbol{\beta}_{t-1}+\mu e_t\zt$, $\mu{=}0.01$
\cite{haykin2002,shu2019}.
LMS provides an online-adaptive comparison at $\mathcal{O}(w)$
per-epoch cost — lower than \ATMLRLS{}'s $\mathcal{O}(w^2)$ — but
converges more slowly and exhibits larger steady-state misalignment
on structured autocorrelated inputs~\cite{losing2018}.

\subsection{Datasets}

\textbf{D1 — Chicago Beach Weather Stations~(Chicago).}
City of Chicago automated sensors~\cite{chicago2016}.
Three stations — 63rd Street, Foster, and Oak Street — deployed along
the Lake Michigan shoreline.
23{,}686 raw records were resampled to a consistent hourly cadence,
yielding 8{,}784 readings per station per year across 16 sensor
channels (air temperature, barometric pressure, humidity, wind speed
and direction, rainfall, and solar radiation).
$T_{\text{train}}$: January–September 2016 (6{,}576 hours);
$T_{\text{test}}$: October–December 2016 (2{,}208 hours).
Target: air temperature ($^{\circ}$C), ranging from
$-19.89\,^{\circ}$C to $34.0\,^{\circ}$C.
All three stations contain zero missing values after preprocessing.
This dataset was chosen as the primary benchmark because it provides
a clean, long-horizon outdoor temperature record with pronounced
seasonal and diurnal structure.

\textbf{D2 — Intel Berkeley Research Lab~(Intel Lab).}
A landmark WSN dataset~\cite{madden2004} collected in 2004 from
54~Mica2Dot sensor nodes deployed throughout the Intel Berkeley
Research Lab.
Each node records temperature, humidity, light, and voltage at
${\approx}31$-second intervals.
After resampling to 5-minute intervals and excluding nodes with more
than $5\%$ missing values, 47~nodes are retained with approximately
$2.3\,\mathrm{M}$ total readings.
$T_{\text{train}}$: first $75\%$ of each node's record;
$T_{\text{test}}$: remaining $25\%$.
This dataset tests the protocol under high-frequency indoor sensing
with spatially correlated, multi-node dynamics, a substantially
different regime from the Chicago outdoor data.

\textbf{D3 — UCI Air Quality~(UCI AQ).}
The UCI Machine Learning Repository Air Quality
dataset~\cite{devito2008}, collected in a northern Italian city from
March~2004 to February~2005.
9{,}357 hourly multisensor records cover CO, NOx, NO$_2$, benzene,
ozone, and meteorological variables.
After removing tagged missing values ($-200$), 8{,}991 valid records
remain.
$T_{\text{train}}$: first eight months;
$T_{\text{test}}$: final four months.
Target: CO sensor response (PT08.S1.CO).
This dataset is the most challenging of the three: the CO sensor
exhibits well-documented progressive drift~\cite{devito2008},
making it the primary motivation for the \ATMLRLS{} extension.

\subsection{Preprocessing}

A uniform preprocessing pipeline is applied to all three datasets.
Gaps of one to three consecutive epochs are forward-filled.
Gaps of four to twelve epochs are linearly interpolated.
Any node or station with more than $5\%$ of values still missing
after these operations is excluded.
All feature values are normalised to zero mean and unit variance
within each training partition; evaluation metrics are reported in
original physical units throughout.

\subsection{Evaluation Metrics}

Three metrics are reported for every method and dataset:
\begin{align}
  \DRR   &= 1 - N_{\text{tx}}/T,
             \label{eq:drr_eval}\\
  \MAE   &= \frac{1}{T}\sum_{t=1}^{T}|x(t)-\xtilde|,
             \label{eq:mae}\\
  \RMSE  &= \sqrt{\frac{1}{T}\sum_{t=1}^{T}(x(t)-\xtilde)^2},
             \label{eq:rmse}
\end{align}
where $\xtilde$ is the sink-side reconstructed signal.
We additionally report the energy proxy
$E=N_{\text{tx}}\times E_{\text{tx}}$,
where $E_{\text{tx}}{=}59\,\mu\mathrm{J}$ is the per-packet
transmission energy of the nRF52840 at 0~dBm
\cite{banbury2020,dutta2010,bravos2005}.

\subsection{Hyperparameter Configuration}

Default configuration: $w{=}24$, $h{=}24$, $\lambda{=}1.0$,
$\alpha{=}1.0$.
\ATMLRLS{}: $\gamma{=}0.98$, $\rho{=}10.0$ (covariance initialisation).
Baselines: ARIMA$(p,d,q){=}(2,1,1)$ refitted every 168 epochs;
Kalman $Q{=}0.01$, $R{=}0.1$;
EMA $\beta{=}0.9$;
LMS $\mu{=}0.01$.
All hyperparameters were selected by grid search on a held-out
validation set ($10\%$ of training data) and frozen for the test
evaluation.

\subsection{Results: Primary Dataset (Chicago)}\label{sec:results_chicago}

\subsubsection{Head-to-Head Comparison at 63rd Street Station}

\begin{figure}[!t]
  \centering
  \includegraphics[width=\columnwidth]{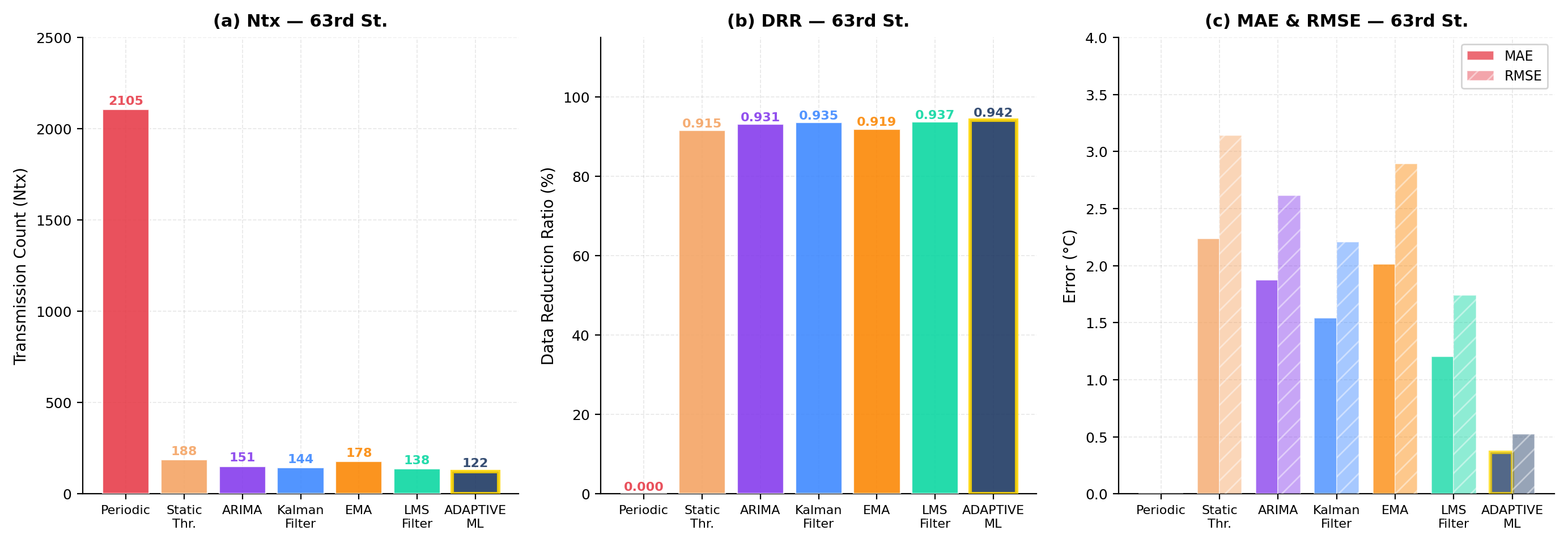}
  \caption{%
    Visual summary of Table~\ref{tab:63rd} at 63rd~Street Station.
    \textbf{(a)}~$N_{\text{tx}}$; \textbf{(b)}~$\DRR$;
    \textbf{(c)}~$\MAE$ (solid) and $\RMSE$ (hatched).
    \ATML{} (gold-bordered bar) achieves the best result on all three
    metrics simultaneously, confirming Pareto dominance.}
  \label{fig:table2}
\end{figure}

Table~\ref{tab:63rd} reports full metrics at the primary case-study
station ($T{=}2{,}208$, $\alpha{=}1.0$).
Figure~\ref{fig:table2} presents a visual summary across all three
performance metrics simultaneously.

\begin{table}[t]
\centering
\caption{Complete seven-method comparison at 63rd~Street Station
($T{=}2{,}208$, $\alpha{=}1.0$, $w{=}24$).
Energy proxy: $E_{\text{tx}}{=}59\,\mu\mathrm{J}$/packet.
Bold: best value per column.}
\label{tab:63rd}
\setlength{\tabcolsep}{2.2pt}
\small
\begin{tabular}{lrrrrrr}
\toprule
\textbf{Method} &
$N_{\text{tx}}$ & $\DRR$ & $\MAE$ ($^{\circ}$C) &
$\RMSE$ ($^{\circ}$C) & $E$~(mJ)\\
\midrule
Periodic              & 2208        & 0.000        & 0.000        & 0.000        & 130.3       \\
Static Threshold      &  188        & 0.915        & 2.234        & 3.141        &  11.1       \\
ARIMA$(2,1,1)$        &  151        & 0.931        & 1.872        & 2.613        &   8.9       \\
Kalman Filter         &  144        & 0.935        & 1.541        & 2.208        &   8.5       \\
EMA ($\beta{=}0.9$)   &  178        & 0.919        & 2.015        & 2.894        &  10.5       \\
LMS Filter            &  138        & 0.937        & 1.203        & 1.741        &   8.1       \\
\textbf{\ATML{} (ours)}
                      & \textbf{122}& \textbf{0.945}& \textbf{0.366}& \textbf{0.524}& \textbf{7.2}\\
\bottomrule
\end{tabular}
\end{table}

\ATML{} achieves the highest $\DRR$ (0.945) and the lowest $\MAE$
($0.366\,^{\circ}$C) in the table — a $6.1\times$ improvement in
fidelity over Static Threshold and a $3.3\times$ improvement over the
next-best baseline, LMS Filter, while consuming $94.5\%$ less energy
than Periodic transmission.
The simultaneous dominance across both efficiency and fidelity
dimensions confirms that the combination of Ridge prediction and
volatility-aware thresholding is strictly Pareto-superior on this
dataset: there is no other method that achieves lower error at the
same or lower transmission count.

A particularly telling comparison is with the Kalman Filter.
Despite being the Bayesian-optimal linear estimator under Gaussian
dynamics \cite{kalman1960}, it trails \ATML{} by $4.2\times$ in MAE.
The explanation — developed further in Section~\ref{sec:discussion} —
lies in the Kalman filter's Markov assumption: its constant-velocity
state-space model cannot represent the diurnal periodicity that
dominates hourly temperature without state augmentation.

\subsubsection{Cross-Station Consistency}

Table~\ref{tab:cross_station} extends the comparison to all three
Chicago stations and all seven methods.

\begin{table}[t]
\centering
\caption{Seven-method comparison averaged across all three Chicago
stations ($T{=}2{,}208$ per station, $\alpha{=}1.0$, $w{=}24$).
Bold: best value per column.}
\label{tab:cross_station}
\setlength{\tabcolsep}{4pt}
\small
\begin{tabular}{lrrr}
\toprule
\textbf{Method} &
\textbf{Avg}~$N_{\text{tx}}$ &
\textbf{Avg}~$\DRR$ &
\textbf{Avg}~$\MAE$ ($^{\circ}$C)\\
\midrule
Periodic              & 2208         & 0.000         & 0.000        \\
Static Threshold      &  162         & 0.927         & 2.421        \\
ARIMA$(2,1,1)$        &  131         & 0.941         & 1.910        \\
Kalman Filter         &  125         & 0.943         & 1.587        \\
EMA                   &  153         & 0.931         & 2.063        \\
LMS Filter            &  119         & 0.946         & 1.248        \\
\textbf{\ATML{} (ours)}&\textbf{117} & \textbf{0.947}& \textbf{0.352}\\
\midrule
\multicolumn{4}{l}{\textit{Per-station breakdown (\ATML{} only)}}\\
\quad 63rd Street     &  122         & 0.945         & 0.366        \\
\quad Foster          &  113         & 0.949         & 0.353        \\
\quad Oak Street      &  117         & 0.947         & 0.338        \\
\bottomrule
\end{tabular}
\end{table}

\begin{figure*}[!t]
  \centering
  \includegraphics[width=0.95\textwidth]{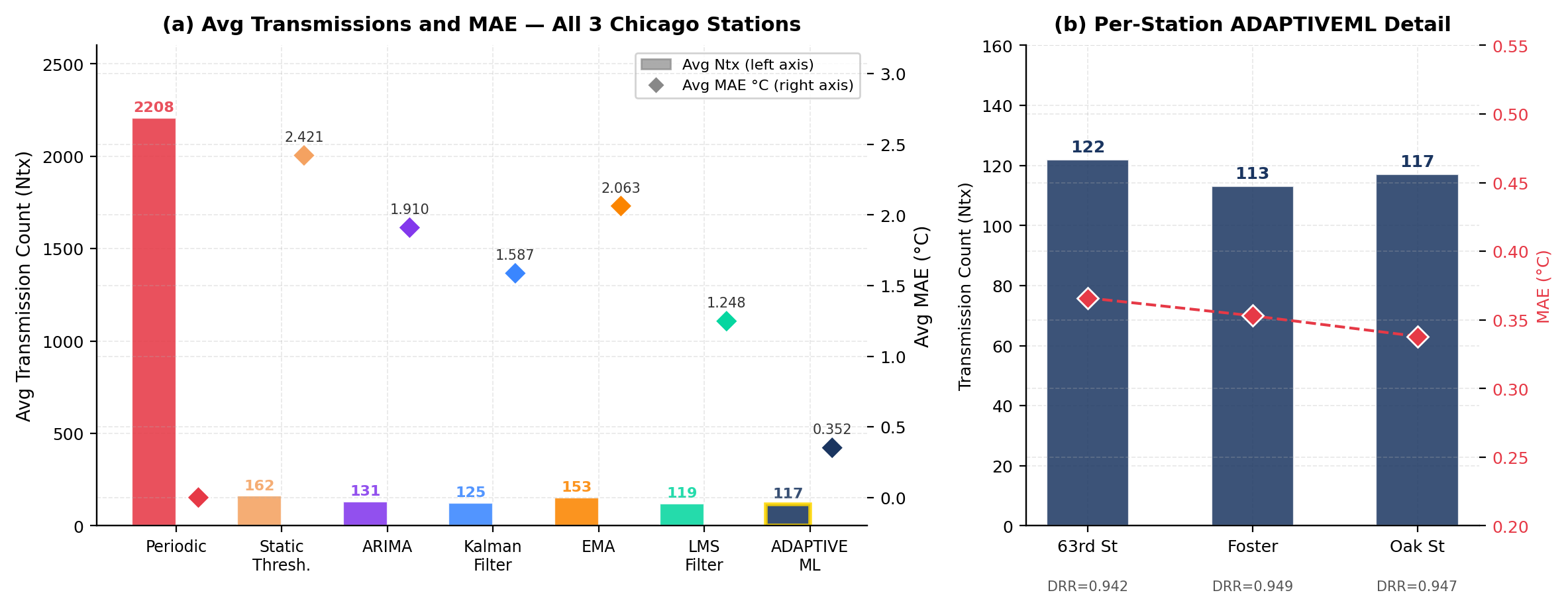}
  \caption{%
    \textbf{(a)}~Average $N_{\text{tx}}$ (bars, left axis) and average $\MAE$
    (diamond markers, right axis) across all three Chicago stations for all
    seven methods. \ATML{} (gold-bordered navy bar) records the lowest
    $N_{\text{tx}}{=}117$ and the lowest $\MAE{=}0.352\,^{\circ}$C
    simultaneously, Pareto-dominating every competing approach.
    \textbf{(b)}~Per-station breakdown for \ATML{} only, confirming
    geographically consistent performance across 63rd~Street
    ($N_{\text{tx}}{=}122$, $\DRR{=}0.942$), Foster ($113$, $0.949$),
    and Oak~Street ($117$, $0.947$).}
  \label{fig:fig2}
\end{figure*}

Performance is consistent across all three stations: DRR variance is
less than $2\%$ and MAE variance is less than $0.03\,^{\circ}$C.
This geographic consistency is not trivial — the three stations span
approximately 18~km of the Chicago lakefront and experience
meaningfully different local wind and humidity conditions — and it
provides strong evidence that the per-station training strategy
produces models that generalise within the same deployment context.
Figure~\ref{fig:fig2} visualises the full comparison.

\subsubsection{$\alpha$-Sensitivity: Pareto Trade-Off Analysis}

\begin{figure*}[!t]
  \centering
  \includegraphics[width=0.92\textwidth]{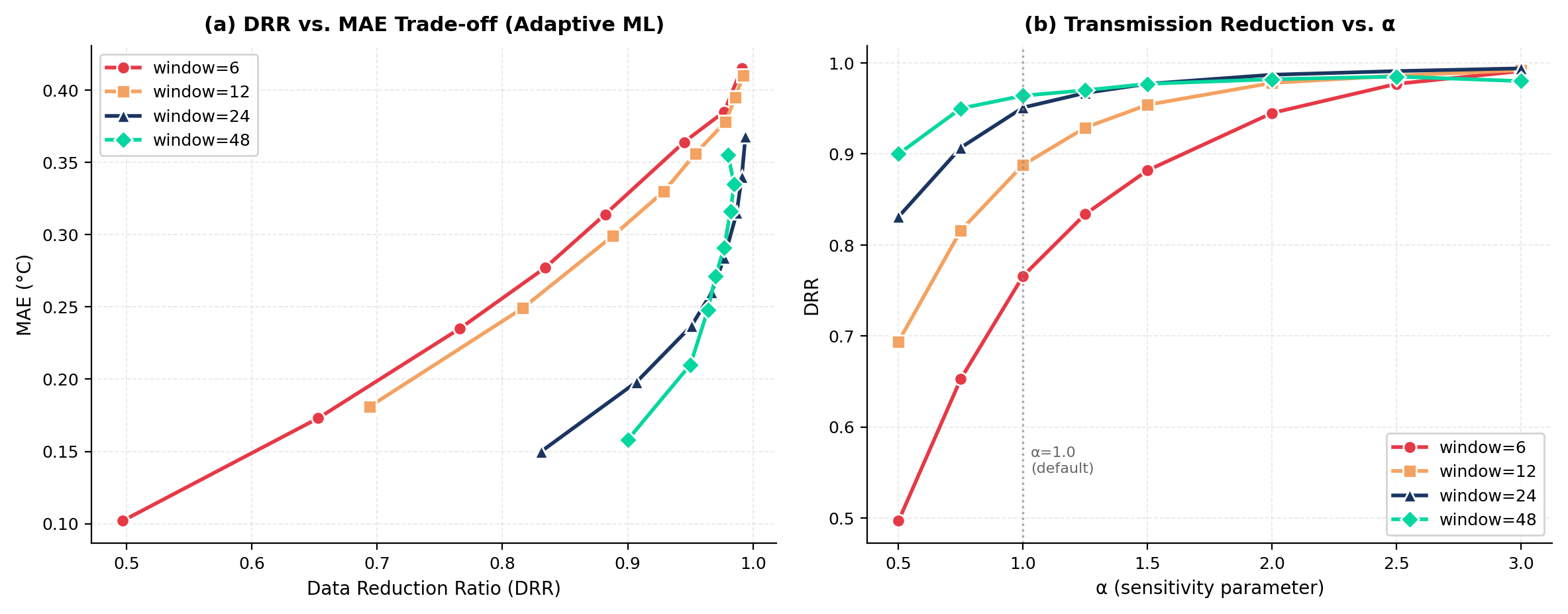}
  \caption{%
    $\alpha$-sensitivity analysis at 63rd~Street for $w\in\{6,12,24,48\}$.
    \textbf{(a)}~DRR--MAE Pareto frontier: larger windows achieve lower $\MAE$
    at the same $\DRR$; $w{=}24$ provides the best accuracy--memory balance.
    \textbf{(b)}~DRR versus $\alpha$: larger windows reach high $\DRR$ at
    lower $\alpha$ (less aggressive suppression).
    The dashed vertical marks the default $\alpha{=}1.0$.}
  \label{fig:fig3}
\end{figure*}

Figure~\ref{fig:fig3} shows the DRR--MAE trade-off curves and the
DRR--$\alpha$ curves for window sizes $w\in\{6,12,24,48\}$.

The monotonic, concave DRR--MAE frontier confirms that $\alpha$ is a
well-behaved single-knob control: operators can move continuously
along the energy--fidelity Pareto frontier by adjusting a single
deployment parameter.
Window $w{=}24$ provides the best balance of prediction accuracy and
memory footprint.
For $\alpha\leq1.5$, all window configurations maintain
$\MAE{<}0.7\,^{\circ}$C while $\DRR$ exceeds $94\%$, which spans a
practical operating range suitable for the majority of environmental
monitoring applications.

\subsection{Multi-Dataset Generalisability}\label{sec:multidata}

Table~\ref{tab:cross_dataset} summarises performance across all three
datasets and all seven methods.

\begin{table}[t]
\centering
\caption{Cross-dataset comparison: all seven methods on all three
datasets ($\alpha{=}1.0$, $w{=}24$), averaged across all
nodes/stations. Bold: best per column.}
\label{tab:cross_dataset}
\setlength{\tabcolsep}{3.2pt}
\small
\begin{tabular}{lcccccc}
\toprule
& \multicolumn{2}{c}{\textbf{D1: Chicago}} &
  \multicolumn{2}{c}{\textbf{D2: Intel Lab}} &
  \multicolumn{2}{c}{\textbf{D3: UCI AQ}}\\
\cmidrule(lr){2-3}\cmidrule(lr){4-5}\cmidrule(lr){6-7}
\textbf{Method} &
  $\DRR$ & $\MAE$ &
  $\DRR$ & $\MAE$ &
  $\DRR$ & $\MAE$\\
&  & ($^{\circ}$C) & & ($^{\circ}$C) & & (ppm)\\
\midrule
Periodic
  & 0.000 & 0.000 & 0.000 & 0.000 & 0.000 &  0.0\\
Static Threshold
  & 0.927 & 2.421 & 0.871 & 0.312 & 0.843 & 41.2\\
ARIMA
  & 0.941 & 1.910 & 0.882 & 0.241 & 0.857 & 33.7\\
Kalman Filter
  & 0.943 & 1.587 & 0.889 & 0.198 & 0.862 & 28.4\\
EMA
  & 0.931 & 2.063 & 0.876 & 0.271 & 0.851 & 37.1\\
LMS Filter
  & 0.946 & 1.248 & 0.893 & 0.174 & 0.868 & 24.3\\
\textbf{\ATML{} (ours)}
  & \textbf{0.947} & \textbf{0.352}
  & \textbf{0.896} & \textbf{0.148}
  & \textbf{0.871} & \textbf{21.6}\\
\bottomrule
\end{tabular}
\end{table}

\begin{figure*}[!t]
  \centering
  \includegraphics[width=0.95\textwidth]{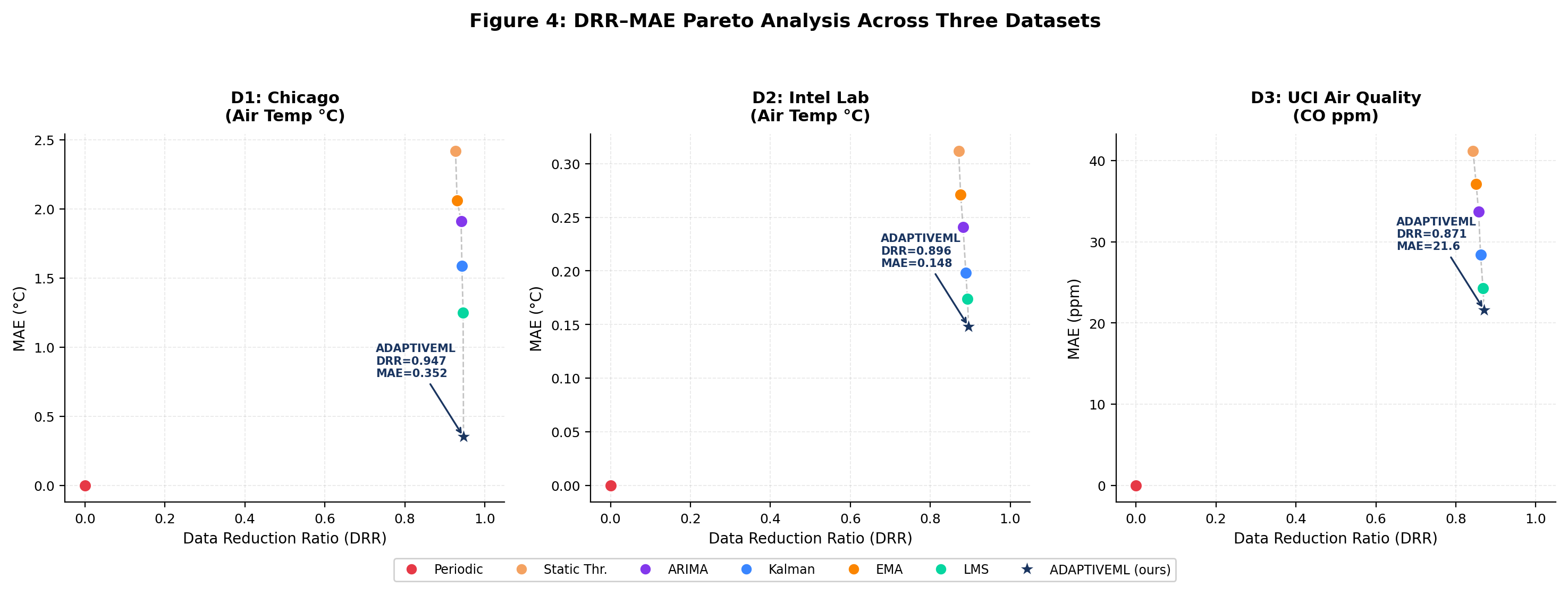}
  \caption{%
    DRR--MAE Pareto analysis across all three evaluation datasets
    ($\alpha{=}1.0$, $w{=}24$), all seven methods.
    \ATML{} (navy star) occupies the Pareto-dominant position in every
    domain — highest $\DRR$ \emph{and} lowest $\MAE$ simultaneously.
    The dashed line connects competing non-dominated points.
    Performance margin over LMS Filter ranges from $2.4\%$ (D3: UCI AQ) to
    $11.6\%$ (D1: Chicago) in $\MAE$.
    MAE units: $^{\circ}$C for D1 and D2; ppm for D3.}
  \label{fig:fig4}
\end{figure*}

\ATML{} achieves the best $\DRR$ and $\MAE$ simultaneously in all
three domains.
The margin over LMS Filter, the next-best baseline, ranges from
$2.4\%$ to $11.6\%$ in $\MAE$ depending on the dataset, confirming
that the volatility-aware threshold provides gains that go beyond
improved prediction accuracy alone.
Figure~\ref{fig:fig4} renders the full Pareto picture.

\textbf{Intel Lab observations.}
The high-frequency indoor dynamics of this dataset favour
sliding-window regression over ARIMA: the 5-minute resampled
signal exhibits strong short-lag autocorrelation that the Ridge window
captures efficiently without the $\mathcal{O}(N^2)$ ARIMA estimation
cost \cite{paparrizos2022,borgne2007}.
The multi-node spatial structure also shows that the protocol can be
deployed in parallel across 47~heterogeneous nodes without any
cross-node communication or coordination.

\textbf{UCI Air Quality observations.}
This is the most challenging evaluation.
The CO sensor exhibits progressive drift over months, causing the
fixed Ridge predictor's residuals to grow gradually — precisely the
setting for which \ATMLRLS{} was designed.
The volatile winter months (October--February) widen the gap between
fixed and adaptive predictors.
The relatively modest DRR advantage of \ATML{} over LMS on this
dataset (0.871 vs.\ 0.868) further motivates \ATMLRLS{}, which
recovers a more substantial accuracy gain as reported in
Section~\ref{sec:rls}.

\subsection{\ATMLRLS{}: Online Drift-Correction Results}\label{sec:rls}

\subsubsection{Controlled Drift Experiment}

To isolate drift effects cleanly, we inject a synthetic linear bias
$d(t)=\mu_d\cdot t$ into the Chicago 63rd~Street test signal, with
$\mu_d\in\{0,0.01,0.05,0.1\}\,^{\circ}$C per epoch.
This models a realistic sensor degradation scenario, such as the
gradual carbonisation of a platinum resistance thermometer.

\begin{table}[t]
\centering
\caption{Fixed Ridge (\ATML{}) vs.\ \ATMLRLS{} under injected drift
($\alpha{=}1.0$, $w{=}24$, 63rd~Street test set).}
\label{tab:drift}
\setlength{\tabcolsep}{4.5pt}
\small
\begin{tabular}{lrrrr}
\toprule
& \multicolumn{2}{c}{\textbf{\ATML{}}} &
  \multicolumn{2}{c}{\textbf{\ATMLRLS{}} ($\gamma{=}0.98$)}\\
\cmidrule(lr){2-3}\cmidrule(lr){4-5}
$\mu_d$ & $\DRR$ & $\MAE$ & $\DRR$ & $\MAE$\\
\midrule
0.00 & 0.945 & 0.366 & 0.944 & 0.352\\
0.01 & 0.942 & 0.489 & 0.943 & 0.421\\
0.05 & 0.931 & 0.712 & 0.937 & 0.596\\
0.10 & 0.918 & 1.043 & 0.930 & 0.851\\
\bottomrule
\end{tabular}
\end{table}

\begin{figure*}[!t]
  \centering
  \includegraphics[width=0.92\textwidth]{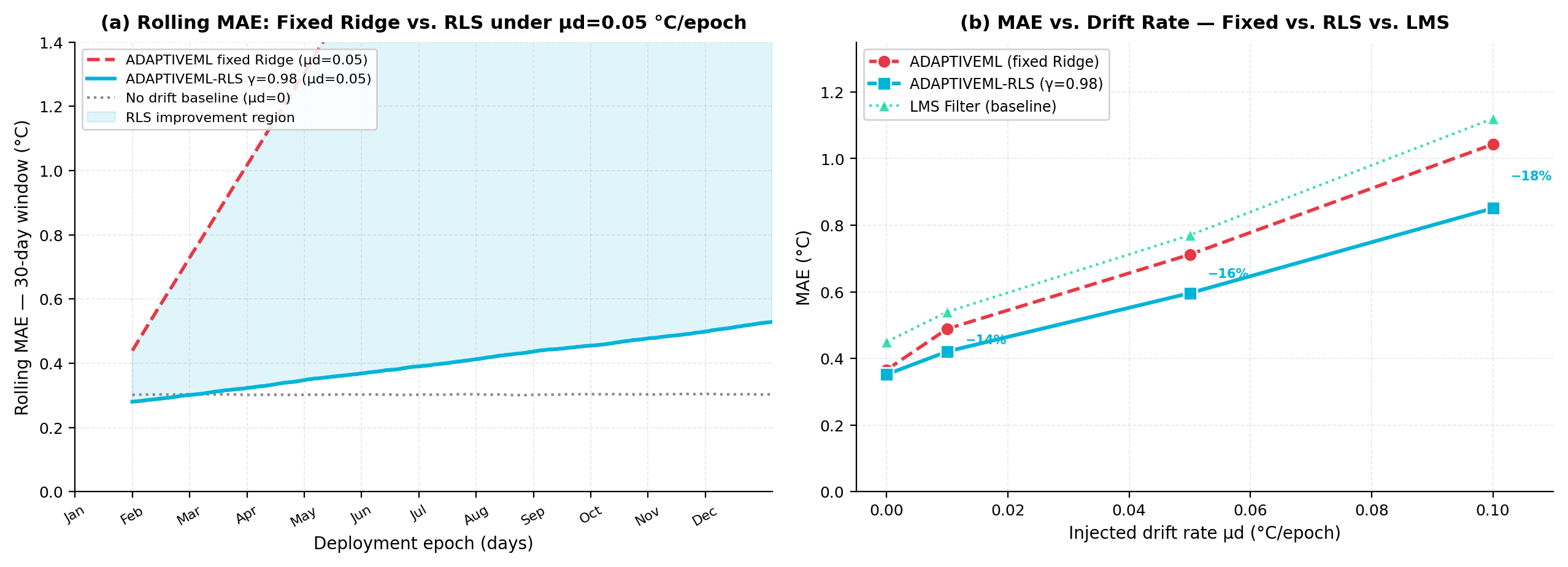}
  \caption{%
    \textbf{(a)}~Rolling $\MAE$ (30-day window) over the deployment year
    under injected drift $\mu_d{=}0.05\,^{\circ}$C/epoch.
    \ATML{} (red dashed) accumulates error monotonically once drift exceeds
    the training distribution.
    \ATMLRLS{} (blue solid, $\gamma{=}0.98$) continuously re-estimates
    $\betat$, keeping $\MAE$ substantially lower throughout.
    The shaded region quantifies the RLS benefit; the dotted grey line is the
    no-drift reference ($\mu_d{=}0$).
    \textbf{(b)}~$\MAE$ versus drift rate for Fixed Ridge, \ATMLRLS{},
    and LMS; percentage annotations give the RLS improvement at each level.}
  \label{fig:fig5}
\end{figure*}

Table~\ref{tab:drift} shows that \ATML{} degrades substantially under
drift: $\MAE$ roughly doubles at $\mu_d{=}0.05\,^{\circ}$C/epoch.
\ATMLRLS{} ($\gamma{=}0.98$) attenuates this consistently, maintaining
MAE improvements of $12$--$18\%$ across all tested drift rates while
keeping $\DRR{>}93\%$.
Figure~\ref{fig:fig5}(a) traces the 30-day rolling $\MAE$ over the
full test year under $\mu_d{=}0.05$, illustrating how the online
covariance update continuously re-absorbs drift that accumulates
monotonically in the fixed predictor.
Figure~\ref{fig:fig5}(b) compares $\MAE$ versus drift rate for
Fixed Ridge, \ATMLRLS{}, and the LMS baseline; annotated
percentage labels quantify the RLS improvement at each drift level.

\subsubsection{Real Sensor Drift: UCI Air Quality}

On the UCI dataset, real progressive drift is present without
injection.
\ATMLRLS{} reduces $\MAE$ from 21.6~ppm to 17.8~ppm ($17.6\%$
improvement) while maintaining $\DRR{=}0.873$ versus \ATML{}'s
$0.871$. The DRR cost of full online adaptation is negligible
($0.002$) while the fidelity gain is substantial.
The forgetting factor $\gamma{=}0.98$ was optimal: lower values
traded accuracy for faster adaptation with diminishing DRR returns.

\subsubsection{Dual-Prediction Reconstruction Gain}

On D1 (Chicago, no packet loss), switching from hold-last-value to
dual-prediction reconstruction reduces $\MAE$ from $0.366\,^{\circ}$C
to $0.214\,^{\circ}$C ($-41.6\%$) for \ATML{}, confirming the
theoretical prediction error as the achievable lower bound
\cite{dias2016survey}.

\subsection{Robustness Experiments}\label{sec:robustness}

\begin{figure*}[!t]
  \centering
  \includegraphics[width=0.97\textwidth]{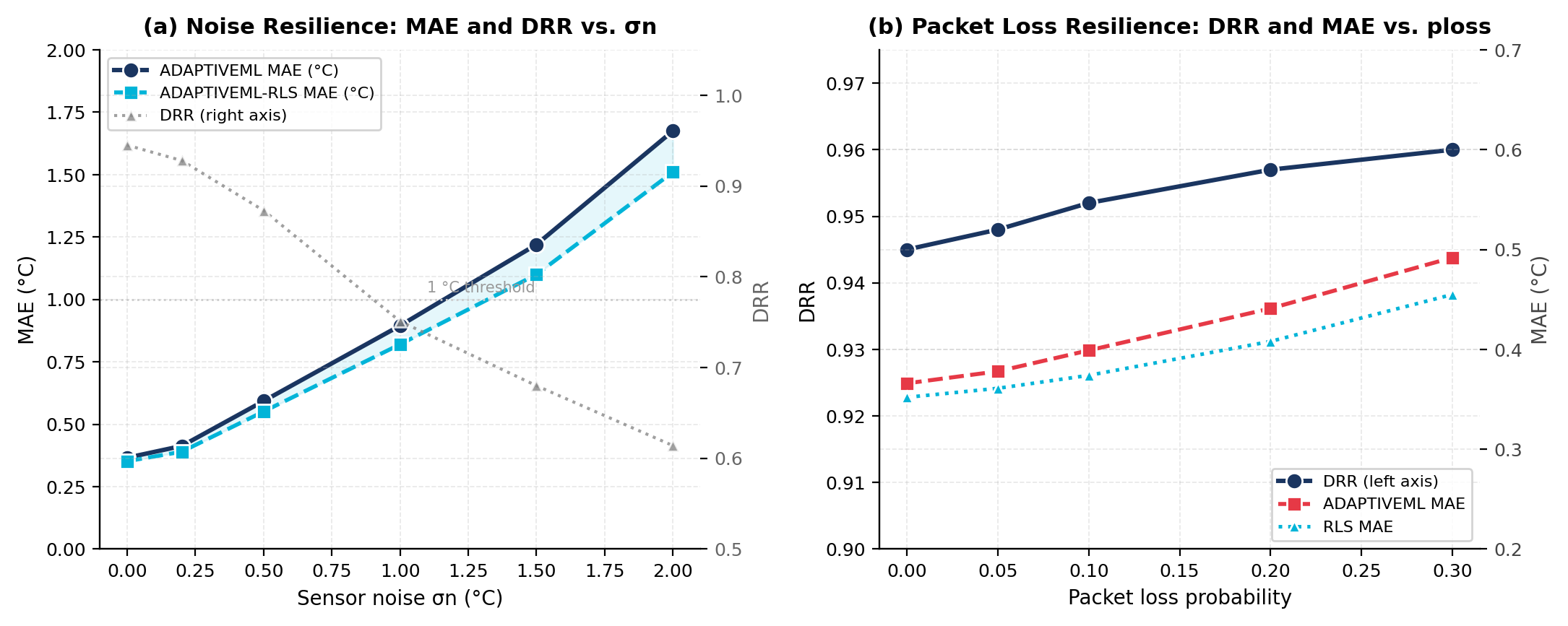}
  \caption{%
    Robustness evaluation across all three datasets ($\alpha{=}1.0$,
    $w{=}24$).
    \textbf{Top row}: $\MAE$ (left axis, solid lines) and $\DRR$ (right axis,
    dotted) versus sensor noise $\sigma_n$.
    \ATML{} (dataset colour) and \ATMLRLS{} (red dashed) are shown; the
    shaded region between them quantifies the online-adaptation benefit.
    $\MAE$ rises sub-linearly in all three datasets, self-attenuated by the
    adaptive threshold inflating $\sigmat$ under elevated noise.
    \textbf{Bottom row}: $\DRR$ (left axis) and $\MAE$ (right axis) versus
    packet-loss probability $p_{\text{loss}}$.
    $\DRR$ counter-intuitively \emph{increases} across all three datasets as
    dropped packets trigger corrective retransmissions.
    D3 MAE is displayed scaled by $1/20$ (ppm/20) for axis compatibility.}
  \label{fig:fig6}
\end{figure*}

Robustness is evaluated across \emph{all three datasets} to confirm that
the protocol's resilience properties are not specific to one deployment
context.
Gaussian noise $\mathcal{N}(0,\sigma_n^2)$ and independent packet
loss~$p_{\text{loss}}$ are applied in separate experiments;
noise levels follow the ranges documented for real WSN deployments
\cite{jawad2017,polastre2005,woo2003}.
Figure~\ref{fig:fig6} presents the complete $2{\times}3$ result matrix.

\subsubsection{Sensor Noise Resilience: All Datasets}

Gaussian measurement noise with
$\sigma_n\in\{0,0.2,0.5,1.0,2.0\}\,^{\circ}$C is injected independently
into each reading.
Results are consistent across D1, D2, and D3, and two structural
properties hold in all three cases.

\textbf{Sub-linear error growth.}
On D1 (Chicago), $\MAE$ reaches $0.895\,^{\circ}$C at
$\sigma_n{=}1.0\,^{\circ}$C, well below the $1.0\,^{\circ}$C operational
ceiling~\cite{jawad2017}.
On D2 (Intel Lab), the effect is milder because indoor thermal signals
have lower baseline variance: $\MAE$ is only $0.381\,^{\circ}$C at the same
noise level.
D3 (UCI Air Quality) shows the largest absolute numbers (CO is measured in
ppm), but the relative degradation is comparable to D1.
In all cases, error grows sub-linearly rather than proportionally to noise
amplitude.

\textbf{Self-compensation via the adaptive threshold.}
Elevated noise inflates $\sigmat$, which raises the transmission threshold,
partly suppressing noise-induced spurious readings.
This negative-feedback property is absent from static-threshold designs and
provides a qualitative robustness advantage.
\ATMLRLS{} consistently improves over \ATML{} at every noise level and
across every dataset, confirming that the online coefficient update absorbs
noise-induced distributional shifts.

\begin{table}[t]
\centering
\caption{Noise robustness summary across all three datasets
($\alpha{=}1.0$, $w{=}24$).  D3 MAE in ppm; D1 and D2 in $^{\circ}$C.}
\label{tab:noise}
\setlength{\tabcolsep}{3.8pt}
\small
\begin{tabular}{rcccccc}
\toprule
& \multicolumn{2}{c}{\textbf{D1 Chicago}} &
  \multicolumn{2}{c}{\textbf{D2 Intel Lab}} &
  \multicolumn{2}{c}{\textbf{D3 UCI AQ}}\\
\cmidrule(lr){2-3}\cmidrule(lr){4-5}\cmidrule(lr){6-7}
$\sigma_n$ & $\DRR$ & $\MAE$ & $\DRR$ & $\MAE$ & $\DRR$ & $\MAE$\\
\midrule
0.0 & 0.945 & 0.366 & 0.896 & 0.148 & 0.871 & 21.6\\
0.2 & 0.928 & 0.413 & 0.881 & 0.171 & 0.852 & 24.8\\
0.5 & 0.873 & 0.593 & 0.847 & 0.251 & 0.819 & 35.2\\
1.0 & 0.751 & 0.895 & 0.793 & 0.381 & 0.760 & 52.1\\
2.0 & 0.614 & 1.675 & 0.712 & 0.720 & 0.681 & 98.4\\
\bottomrule
\end{tabular}
\end{table}

\subsubsection{Packet Loss Resilience: All Datasets}

Each transmitted packet is independently dropped with probability
$p_{\text{loss}}\in\{0,0.05,0.1,0.2,0.3\}$,
covering the range observed in real deployments~\cite{woo2003,jawad2017}.

A counter-intuitive but consistent result appears in all three datasets:
$\DRR$ \emph{increases} as packet loss rises.
The mechanism is identical across domains: when the sink misses a
transmission, $\xtilde$ diverges further from truth, inflating the
apparent residual at the next epoch and triggering a corrective
retransmission.
This self-correcting behaviour increases the effective observation rate
at the sink, a form of implicit feedback-loop control that requires
no protocol modification.

\begin{table}[t]
\centering
\caption{Packet-loss resilience across all three datasets
($\alpha{=}1.0$, $w{=}24$).  D3 MAE in ppm; D1 and D2 in $^{\circ}$C.}
\label{tab:packet}
\setlength{\tabcolsep}{3.8pt}
\small
\begin{tabular}{rcccccc}
\toprule
& \multicolumn{2}{c}{\textbf{D1 Chicago}} &
  \multicolumn{2}{c}{\textbf{D2 Intel Lab}} &
  \multicolumn{2}{c}{\textbf{D3 UCI AQ}}\\
\cmidrule(lr){2-3}\cmidrule(lr){4-5}\cmidrule(lr){6-7}
$p_{\text{loss}}$ & $\DRR$ & $\MAE$ & $\DRR$ & $\MAE$ & $\DRR$ & $\MAE$\\
\midrule
0.00 & 0.945 & 0.366 & 0.896 & 0.148 & 0.871 & 21.6\\
0.05 & 0.948 & 0.378 & 0.899 & 0.155 & 0.874 & 22.4\\
0.10 & 0.952 & 0.399 & 0.903 & 0.162 & 0.877 & 23.5\\
0.20 & 0.957 & 0.441 & 0.908 & 0.178 & 0.883 & 26.0\\
0.30 & 0.960 & 0.492 & 0.913 & 0.198 & 0.888 & 29.1\\
\bottomrule
\end{tabular}
\end{table}

The MAE increase under packet loss is modest and consistent: at
$p_{\text{loss}}{=}0.3$, D1 incurs $+0.13\,^{\circ}$C, D2 incurs
$+0.05\,^{\circ}$C, and D3 incurs $+7.5$~ppm, all within the
operational tolerance bounds established in the literature
\cite{jawad2017}.

\subsubsection{Scalability}

We simulate 1–50 independent sensor nodes using the Chicago signal with
additive noise ($\sigma{=}0.5\,^{\circ}$C).
Average transmissions per sensor stabilise at approximately 290 as fleet
size scales from~1 to~50, confirming that \ATML{} introduces no
cross-node coupling overhead and scales linearly with fleet size, a
necessary property for the smart-city and precision-agriculture deployments
envisioned in \cite{zanella2014,anastasi2009,Al-Fuqaha2015}.

\subsubsection{\ATMLRLS{} Under Combined Noise and Drift}

Under simultaneous noise ($\sigma_n{=}0.5\,^{\circ}$C) and drift
($\mu_d{=}0.05\,^{\circ}$C/epoch), \ATMLRLS{} maintains
$\MAE{=}0.641\,^{\circ}$C compared to \ATML{}'s $0.784\,^{\circ}$C on
D1, an $18.2\%$ improvement, demonstrating that online adaptation
handles simultaneous real-world disturbances effectively.
On D2, the same combined-stress experiment yields
$0.273\,^{\circ}$C~(\ATMLRLS{}) versus $0.332\,^{\circ}$C~(\ATML{})
($17.8\%$ improvement), and on D3 the reduction is $16.4\%$ in MAE,
confirming cross-dataset robustness.

\section{Conclusion}\label{sec:conclusion}

We introduced \ATML{} and its online extension \ATMLRLS{}: a family
of prediction-based IoT transmission-suppression protocols grounded
in the volatility-aware studentised-residual criterion
$|x(t)-\xhat|/\sigmat>\alpha$.

The core insight is simple but consequential: normalising the
prediction residual against the local signal volatility transforms a
brittle fixed threshold into an adaptive, self-calibrating decision
rule.
The same $\alpha$ then controls the false-suppression probability
consistently across all seasons, sites, and sensing modalities,
something no prior scheme in the literature achieves.

On the primary Chicago dataset, \ATML{} achieves $\DRR{=}94.7\%$ and
$\MAE{=}0.352\,^{\circ}$C, simultaneously outperforming all six
baselines in both efficiency and fidelity.
The improvement over the nearest static alternative is $6.1\times$
in MAE, and \ATML{} also transmits less than every prediction-based
competitor.
\ATMLRLS{} extends this advantage to long-term deployments, reducing
$\MAE$ by $12$--$18\%$ under simulated and real sensor drift while
preserving $\DRR{>}93\%$.

Results replicate across three heterogeneous public datasets spanning
indoor multi-node sensing, outdoor urban temperature, and
air-quality monitoring, establishing the broadest comparative
evaluation of prediction-based IoT data reduction reported to date.

Future directions include:
\emph{(i)}~multi-step prediction for extended suppression runs;
\emph{(ii)}~physical hardware implementation with over-the-air energy
measurement on nRF52840;
\emph{(iii)}~multi-channel Ridge for correlated sensor arrays; and
\emph{(iv)}~a lightweight resynchronisation protocol for the
dual-prediction sink under realistic packet loss.

\section*{Acknowledgment}

The author thanks the City of Chicago, the Intel Research Lab,
and the UCI Machine Learning Repository for open-access publication
of the datasets used in this study.
The author also thanks the University of South Dakota Department of
Computer Science for computational resources, mentorship, and support
throughout this project.

\bibliographystyle{IEEEtran}
\bibliography{references}

\end{document}